\documentclass{elsart1p}

 \usepackage{graphicx}

\usepackage{amssymb}

\begin{document}

\begin{frontmatter}


 \title{Quark and Gluon Degrees of Freedom in High-Energy Heavy Ion Collisions}

 \author{Rainer J.~Fries}

 \ead{rjfries@comp.tamu.edu}

 \address{Cyclotron Institute, Texas A\&M University, College Station, TX
   77843}
 \address{RIKEN/BNL Research Center, Brookhaven National Laboratory, Upton, NY
   11973}


\begin{abstract}
I discuss some recent progress in our understanding of 
high energy nuclear collisions. I will focus on two topics
which I was lucky to co-pioneer in the recent past. One is recombination of
quarks and its interpretation as a signal for deconfinement, the second
is electromagnetic radiation from jets passing through a quark gluon plasma.
This talk was given during the award ceremony for the 2007 IUPAP Young
Scientist Award.
\end{abstract}

\begin{keyword}
Relativistic Heavy Ion Collisions \sep Quantum Chromodynamics

\PACS 25.75.Dw \sep 24.85.+p
\end{keyword}
\end{frontmatter}

\section{Introduction}

Nuclear collisions at center of mass energies $\sqrt{s_{NN}} \gg 1$ GeV
are carried out to look for new phases of quantum chromodynamics (QCD) 
in which quark and gluon degrees of freedom are explicit. A phase 
transition or rapid cross over into a deconfined quark gluon plasma (QGP) 
is found in lattice QCD calculations around $T_c\approx 180$ MeV and 
energy densities $\epsilon_c \approx 1$ GeV/fm$^3$ \cite{Karsch:2001vs}. 
At the highest 
available energies, achieved at the Relativistic Heavy Ion Collider (RHIC, 
$\sqrt{s_{NN}} = 200$ GeV), we have now convincing evidence that this 
new state of matter has been created \cite{Adams:2005dq}.
This discovery has also come with many a surprise, e.g.\ strong 
indications from data that the matter at RHIC is far from an asymptotically 
free gas of quarks and gluons, but rather strongly coupled 
\cite{Gyulassy:2004zy}.

In these proceedings, I discuss two topics that emerged during the 
past couple of years after RHIC started running in the year 2000. The first
one, recombination of quarks, was driven by experimental results which
contradicted the way hadrons were expected to be created in high energy
collisions, through fragmentation from QCD jets. This was called the baryon 
puzzle or baryon anomaly in the early RHIC years. The solution is 
surprisingly straight forward and can be found in a simple recombination 
or coalescence picture which is valid in a phase space filled with partons. 
This was proposed around the same time in early 2003 by me and my 
collaborators at Duke University as well as by a few other groups. A completely
satisfying dynamical description of this process is still lacking due to 
its non-perturbative nature, but a simple counting 
rule emerging from this picture has some very powerful implications 
as I will discuss below. 

The second topic involves a new class of signals from the hot matter 
created in the collisions. It was long hoped that electromagnetic probes --- 
photons and lepton pairs --- can, due to their penetrating nature, shine light 
on the conditions inside the fireball and at early times during the collision.
On the other hand, QCD jets are used as probes for the opacity and other
transport properties of the medium. We proposed that the two can be combined
in a novel way. Jets traveling through the QGP can produce real and virtual
photons. We found that this process can contribute significantly to the 
total photon yield and that it can be used to learn about the medium. 
Unlike the case of quark recombination which grew out of an 
experimental puzzle and has since then been tremendously successful for the
phenomenology at RHIC, the experimental sensitivity is not yet sufficient to 
routinely use photons from jets. This will change with the
luminosity upgrade for RHIC which will permit the use of more exotic 
and extremely powerful probes.

\section{Quark Recombination}

\begin{figure}[b]
\centering
\includegraphics[width=7cm]{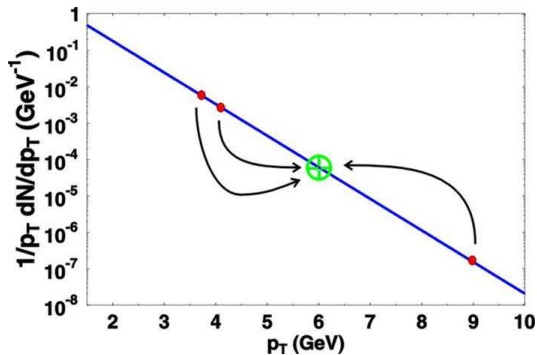}
\caption{\label{fig1} Sketch of the fragmentation and recombination mechanism
 working on an exponential quark spectrum (black line). To create a 6 GeV
 meson (green dot), fragmentation needs to start from a quark or gluon with 
 an average momentum of $\sim$ 9-12 GeV/$c$, which is exponentially suppressed.
 Recombination uses two quarks with roughly 3 GeV/$c$ which is much more 
 likely.}
\end{figure}

We have learned in the past that hadrons produced with transverse 
momenta $P_T$ of more than $\sim 1$ GeV/$c$ come from QCD jets, originating 
from a single quark or gluon with large momentum $p$ which fragments into 
a shower of hadrons. A given hadron has a fraction $z$ of the original 
momentum, $P_T = zp$, $0<z<1$, and the probabilities for the fragmentation
process are universal \cite{Collins:1981uw}. Among other things universality 
predicts a dominance of mesons over baryons. E.g.\ the ratio of protons 
over pions is expected to be roughly $0.2 \ldots 0.3$. However, in nuclear 
collisions at RHIC a ratio of $p/\pi \approx 1$ was found for $P_T \approx 4$ 
GeV/$c$, well in the range where fragmentation was expected to work
\cite{Abelev:2006jr}.

The solution of this puzzle can be found by realizing that fragmentation 
necessitates the absence of any other partons which might interact with 
the jet. This puts tight limits on the applicability of the fragmentation
picture to nuclear collisions with thousands of particles created. 
Rather, with phase space filled with partons in a thermalized medium, 
hadron production should proceed through recombination or
coalescence of quarks into the valence structure of hadrons
\cite{FMNB:03prl,FMNB:03prc,Fries:04qm,GreKoLe:03prl,HwaYa:02}. Most
implementations of the recombination process use an instant projection 
of quark states onto hadron states utilizing the wave function 
$\psi$ of the baryon
or meson and assuming thermal distribution functions $f$ for quarks.
E.g. for pions one has
\begin{equation}
  \frac{dN_\pi^+}{d^3 P} = C_{\pi^+} 
   \int d\Sigma \int \frac{d^3 q}{(2\pi)^3} f_u (P/2-q)
   f_{\bar d} (P/2+q) |\psi(q)|^2 
\end{equation}
where $\Sigma$ is the hypersurface of hadronization and $C_{\pi^+}$ is a
combinatorial factor \cite{FMNB:03prc}. 
This ansatz preserves 3-momentum, but not energy, and it is hence only 
valid for not too small momenta $P_T$.

\begin{figure}[b]
\centering
\includegraphics[width=10cm]{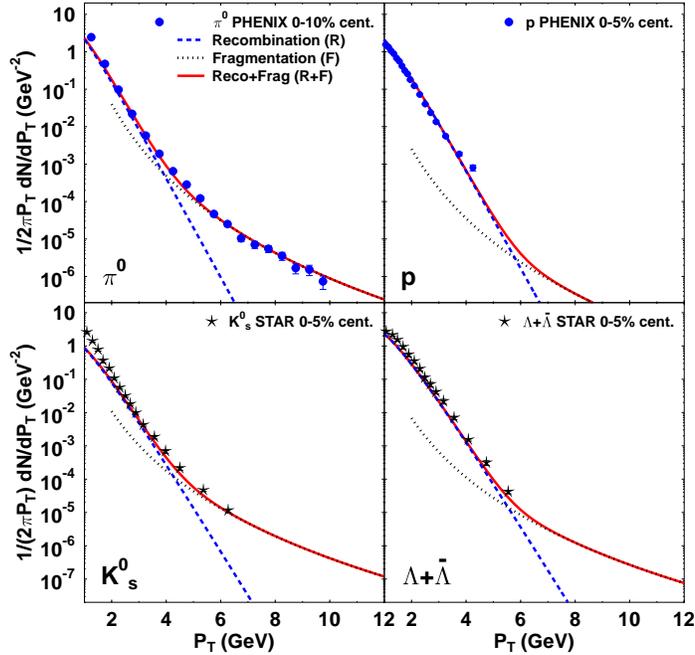}
\caption{\label{fig2} Spectra for pions, protons, kaons and Lambdas in
  a combined recombination and fragmentation approach \cite{FMNB:03prc} 
  compared to data from RHIC. Shown are the fragmentation contribution 
  (dotted line), recombination from the thermal medium with temperature 
  $T_c$ (dashed line) and 
  the sum of both contributions. The different transition regions for
  mesons and baryons are clearly visible.}
\end{figure}

For thermal quark distributions $f\sim e^{-p/T}$ there is no suppression 
of baryon production compared to mesons, leading naturally to baryon/meson 
ratios of order unity. On the other hand, one can show that recombination
on thermal spectra is always more efficient than fragmentation, as
indicated in Fig.\ \ref{fig1}. Fig.\ \ref{fig2} shows a typical result
for $P_T$ spectra of several hadron species compared with experimental 
data from RHIC (see \cite{FMNB:03prc} for details). The transition from the 
recombination dominated domain to the fragmentation dominated region occurs 
around $P_T \approx 4$ GeV/$c$ for mesons. For baryons the transition is 
shifted to about 6 GeV/$c$ due to the inefficiency of the fragmentation 
process for baryons. Below $P_T \approx 1.5$ GeV/$c$ the 
simple projection formula loses its validity.

Recombination has a very intriguing consequence for elliptic flow $v_2$.  
Elliptic flow arises from the ellipsoidal shape of the overlap zone
of the two nuclei for finite impact parameters $b>0$. In those cases there
is no spherical symmetry in the transverse plane. As a
result the pressure gradients along the smaller and larger 
transverse radii of the fireball are different and lead to a larger boost 
of particles in the direction where the fireball was originally thinner.
The final particle spectra can be analyzed in a harmonic series in the 
azimuthal angle $\phi$
\begin{equation}
  1 + 2 v_2 \cos 2\phi + \ldots  \, .
\end{equation}

\begin{figure}
\centering
\includegraphics[width=7cm]{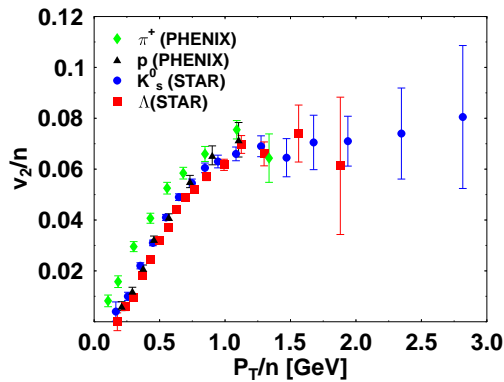}
\caption{\label{fig3} Measured elliptic flow $v_2$ for different hadron
 species at RHIC as a function of $P_T$. The different data sets collapse 
 onto a single line if both axes are scaled by the number of valence quarks
 for each hadron species. Small deviations from universality can be 
 explained in a scaling model using transverse kinetic energy instead of 
 $P_T$.}
\end{figure}

Suppose elliptic flow is born in a partonic phase, and the value for 
quarks just before hadronization is $v_2^p(P_T)$. Recombination then
predicts the value of $v_2$ for all hadron species just after hadronization.
Just using the simplest assumptions about the recombination process
this leads to a universal scaling law
\begin{equation}
  v_2^h (P_T) = n v_2^p (P_T/n)
\end{equation}
where $h$ is any hadron and $n$ is its number of valence quarks
\cite{FMNB:03prc,GreKoLe:03prl,Voloshin:02}.
Hence recombination predicts that the elliptic flow for all meson
species is the same, and that the same is true for all baryons
Moreover it predicts that baryons and mesons are related by a simple 
scaling of both $v_2$ and $P_T$ by 3 and 2 respectively. This is 
impressively confirmed by experimental data \cite{STAR:03v2,PHENIX:03v2}
as shown in Fig.\ \ref{fig3} where $v_2$ for different hadrons in plotted 
on scaled axes. All data
points fall on one universal curve which indicate the quark $v_2$. 
The scaling is even more impressive if the transverse kinetic energy is
used instead of $P_T$, an indication for the hydrodynamic origin of
flow (see \cite{Ravagli:2007xx} for a recent attempt to explain kinetic
energy scaling).

There are many caveats to this simple picture, some of which have 
been successfully overcome, while others are still puzzling.
As an example for the former, let me mention 2-particle correlations 
\cite{STARcorr:02,phenix:04corr} which first seemed to be incompatible 
with recombination, but have now been implemented in a fashion
compatible with data \cite{FMB:04, FB:05}. On the other hand a 
persisting problem is the fact that recombination models are very 
sensitive to space-momentum correlations 
in the quark phase and in connection with data give tight constraints, 
in contradiction to other models \cite{Pratt:2004zq}.

Nevertheless, recombination is an extremely successful model for hadron 
production at intermediate $P_T$ in high energy nuclear collisions. It also
leads to some more general insight. The scaling law in particular makes 
the quark degrees of freedom in hadrons explicit. However, unlike quark
counting rules in elementary processes,
the observable $v_2$ describes the effect of matter moving collectively. 
The collectivity together with the scaling law leaves no doubt that the 
hydrodynamic expansion of the system starts in a phase which is partonic, 
not hadronic. Therefore, the scaling law for $v_2$ might be the best 
\emph{direct} signal for deconfinement available at this moment.

\section{Photons from Jets}

Let us now discuss jets and electromagnetic signals from their interaction
with quark gluon plasma. Photons and dileptons (from virtual photons) have
long been considered as unique probes of dense nuclear matter, since their
mean free path exceeds by far the size of a nucleus. Therefore, photons 
even when emitted deep inside the fireball or very early during the collision 
will reach the detectors unaltered. Thermal radiation of photons and dileptons
is supposed to be the ideal thermometer for the plasma. However, in the 
history of heavy ion collisions, electromagnetic probes have always been
a challenge for experimentalists, requiring large numbers of events and 
mature analysis tools. E.g.\  the background from $\pi^0$ decays is very
problematic for the extraction of direct photons. 

\begin{figure}[b]
\centering
\includegraphics[width=3.25cm]{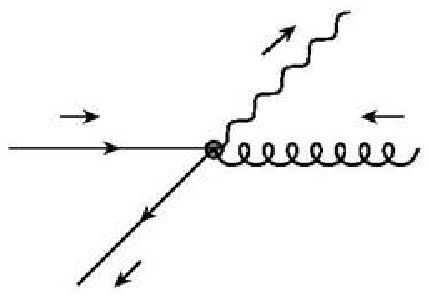}
\includegraphics[width=3.25cm]{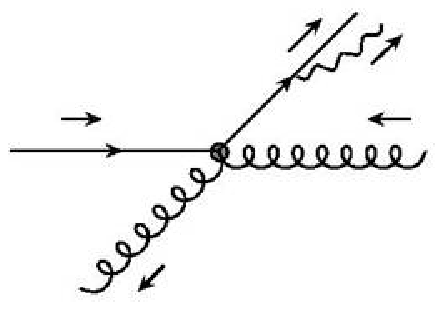}
\includegraphics[width=3.25cm]{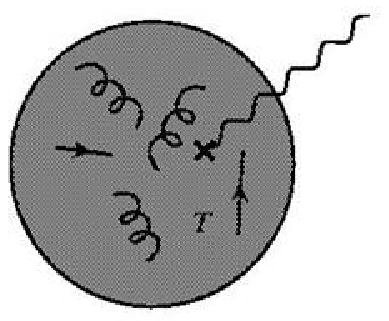}
\includegraphics[width=3.25cm]{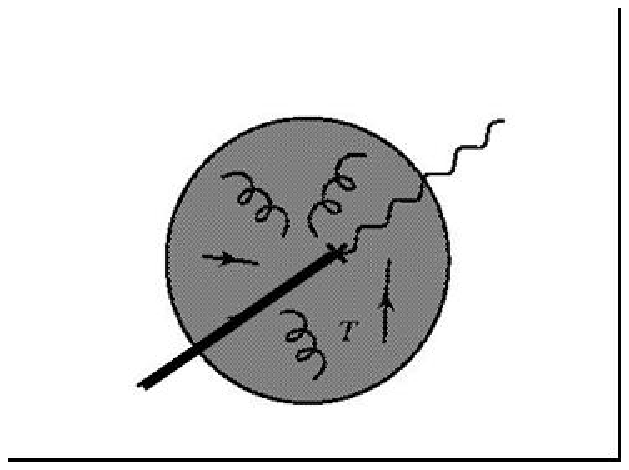}
\caption{\label{fig4} Different contributions to direct photon production
in heavy ion collisions. From left: prompt photons from initial hard 
scatterings of partons; (vacuum) bremsstrahlung from jets; thermal photons
from the quark gluon plasma; jets emitted from jet-plasma interactions}
\end{figure}

Even after subtracting decay photons, there are several sources of direct 
photons besides the thermal radiation, as indicated in Fig.\ \ref{fig4}.
Prompt photons from initial hard scatterings and bremsstrahlung from jets
are present in elementary proton-proton collisions as well as nuclear 
collisions. Thermal radiation appears in nuclear collisions, both from
a hot hadronic medium and from a QGP. $P_T$ as a variable
helps to distinguish the different sources, with the exponential thermal
spectrum being most prominent below 1 GeV/$c$, and with bremsstrahlung and
direct hard photons becoming dominant at large $P_T$. 

In 2002, we suggested that jets traveling through the plasma can radiate 
photons (and dileptons) as well and that this might be an important process
\cite{Fries:2002kt,Gale:2004ud}. 
The leading order channels for real photons are
annihilation, $q$ (jet)$ +\bar q$ (medium) $\to \gamma + g$ and
Compton scattering $q$ (jet)$ +g$ (medium) $\to \gamma + q$. 
These processes creating photon radiation come about naturally 
once the presence of gluon radiation (involved in the energy loss of jets 
\cite{PHENIX:03pi0}) 
has been established.
Since the corresponding cross sections are strongly forward and backward 
peaked, the resulting photon spectra are directly proportional to the
input jet spectra. Therefore, the term jet-photon conversion was coined
for these processes.

\begin{figure}
\centering
\includegraphics[width=8cm]{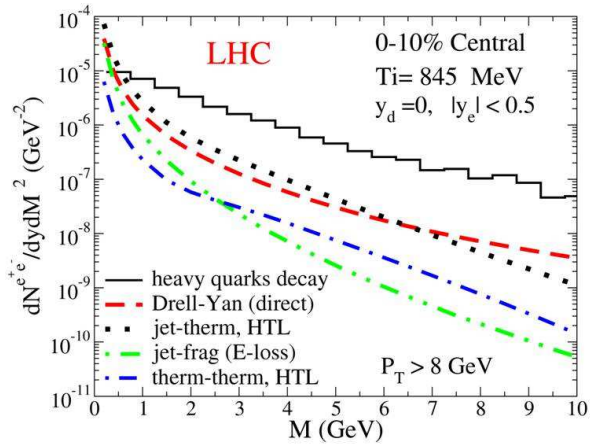}
\caption{\label{fig5} Yield for electron-positron pairs from different sources
  at the LHC \cite{Turbide:2006mc}. The sources are thermal radiation 
  (dash-dotted line), direct primary
  hard scattering (dashed line), bremsstrahlung (dash-dot-dotted line) and 
  jet-medium interactions (dotted line). The background from correlated 
  charm and bottom decays is shown as well. Even for relatively large 
  $P_T>8$ GeV/$c$  jet-medium dileptons are still the dominant source 
  at intermediate masses.}
\end{figure}

Initially we found that the brightness of this new source is comparable
to the other sources at intermediate $P_T$ of a few GeV/$c$.
Refined calculations, also taking into account energy loss of the jet 
before the photon is produced, confirm this result 
\cite{Fries:2005zh,Turbide:2005fk,Turbide:2006mc}. Fig.\ \ref{fig5}
shows a recent calculation for expected dilepton yields at the Large 
Hadron Collider (LHC, $\sqrt{s_{NN}} = 5.5$ TeV) as a function of
dilepton mass $M$ \cite{Turbide:2006mc}. At LHC the relative contribution 
from jet-medium photons will be even larger than at RHIC.

Why should one be excited about this new source of photons? 
Obviously it is important to know all contributions. But there is more,
jet-medium photons are sensitive to the temperature, similar to
thermal radiation, and can be used as a second, independent constraint 
on the temperature evolution of the fireball. Moreover, they can be measured
at a $P_T$ of several GeV/$c$ where the $\pi^0$ background is suppressed by
a factor of 5 due to jet quenching \cite{PHENIX:03pi0}, a luxury which
is not happening at small $P_T$.

They are also sensitive to jet energy loss. However, they probe
jet path integrals different from those which determine hadron observables.
The latter probe the propagation of a jet to the space-time boundary of 
the fireball, while the former only probe up to the point of photon 
production. Thus jet-medium photons also encode information about energy loss
which is complementary to that contained in hadronic observables. 

All of this is very difficult to extract from single inclusive measurement, 
e.g.\ the photon spectrum as a function of $P_T$. In fact, although the 
photon spectra measured at RHIC can be nicely described by calculations 
including jet-medium photons, parameter space is flexible enough to allow
for fits of the data excluding jet-medium interactions. To resolve this issue
one has to go to less inclusive measurements and to correlations.
Therefore, we recently suggested photon elliptic flow as an interesting
observable \cite{Turbide:2005bz}. Photons from thermal emission and photons
from vacuum bremsstrahlung (which experience energy loss) have elliptic
flow $v_2 >0$, i.e.\ in phase with the $v_2$ of hadrons. There are more 
photons from these sources in the direction where the fireball was 
originally thinner. However, the opposite is true for jet-medium photons. 
The thicker the medium, the more likely it is that the jet is converted 
to a photon (remember that this is a rare process). Thus the $v_2$ 
for jet-medium photons should be negative, providing an additional 
possibility to distinguish it from other sources.

\begin{figure}
\centering
\includegraphics[width=10cm]{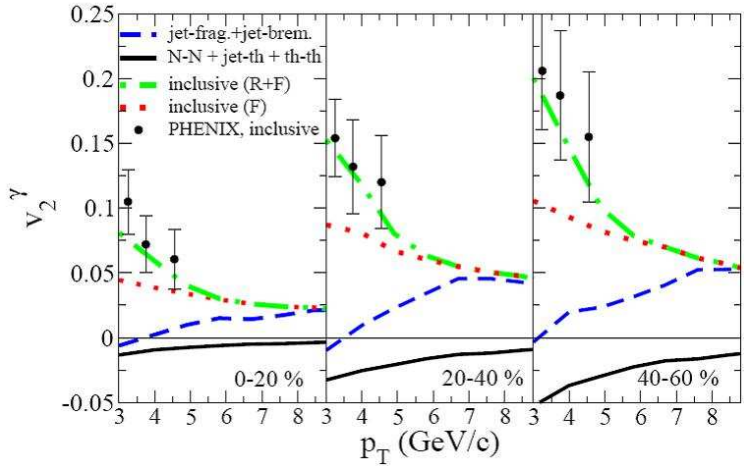}
\caption{\label{fig6} Elliptic flow of photons from different sources
 for three different centralities as a function of photon $P_T$
 \cite{Turbide:2005bz}.
 The result for primary hard photons, thermal photons and jet-medium photons
 (solid line) and bremsstrahlung photons (dashed line) are shown 
 separately. [Bremsstrahlung photons here also include 
 medium-induced bremsstrahlung (besides the vacuum part) which is not 
 discussed in detail in the text. Medium-induced bremsstrahlung can also lead
 to negative $v_2$.]
 The reason to show both results is the hope
 that both contributions might in the future be separable by isolation cuts.
 The effect of negative $v_2$ is clearly visible and can be as large as 
 $-5$\%. 
 The dotted line shows the expected $v_2$ of all photons including decays 
 of pions from fragmentation, the dashed line also includes decays of pions
 from recombination. Data from PHENIX is for inclusive photons without
 background subtraction.}
\end{figure}

Fig.\ \ref{fig6} shows the result of a recent calculation of photon
elliptic flow from \cite{Turbide:2005bz}. The expected $v_2$ of direct 
photons is numerically small and can reach negative values at intermediate 
$P_T$. The data 
shown still contains the background from $\pi^0$ and $\eta$ decays. 
First attempts by PHENIX to extract the $v_2$ of direct photons yielded
results which are compatible with zero with rather large error bars
\cite{Adler:2005rg}.
Future analyses will improve these results and give a definite answer
about the importance of jet-medium photons and dileptons. Even more
promising, but even harder to extract experimentally are jet-photon
and hadron-photon correlations. Such challenging but powerful measurements
will make jet-medium photons an important topic for future runs at RHIC 
and LHC.

\section{Conclusions}

I am particularly grateful to all my collaborators on these projects, 
Steffen A.\ Bass, Charles Gale, Berndt M\"uller, Chiho Nonaka,
Dinesh K.\ Srivastava and Simon Turbide. I wish to thank the 
IUPAP C-12 committee for the honor to be a recipient of the IUPAP 
Young Scientist Award and the organizers of INPC 2007 for a wonderful 
conference experience. This work is supported by DOE grant DE-AC02-98CH10886, 
the RIKEN/BNL Research Center and the Texas A\&M College of Science.

\end{document}